\documentclass[11pt]{article}
\usepackage{graphicx,amsmath,amssymb,epsf}
\usepackage{latexsym,bm, slashed}
\usepackage{xcolor}
\definecolor{dark}{rgb}{0.10,0.2,0.3}
\definecolor{magenta}{rgb}{0.7,0.1,0.3}
\definecolor{purpure}{rgb}{0.5,0.15,0.3}
\usepackage[font=small,format=plain,labelfont=bf,up,textfont=it,up]{caption}
\usepackage{hyperref, cite}
\hypersetup{colorlinks,
citecolor=blue,
filecolor=blue,
linkcolor=magenta,
urlcolor=purpure,hyperfootnotes=true,pdftex}

\title{\bf \Large  BFKL  evolution and the growth with energy of exclusive  $J/\Psi$  and
  $\Upsilon$ photo-production  cross-sections}
\author{ I.~Bautista${}^{1}$, A.~Fernandez~Tellez${}^1$ and  M.~Hentschinski${}^{1,2}$
\bigskip \\
{ $^1$Facultad de Ciencias F\'isico Matem\'aticas, } \\ {Benem\'erita
  Universidad Aut\'onoma de Puebla, Puebla 1152, Mexico}
\\
  {$^2$Instituto de Ciencias Nucleares, Universidad Nacional Aut\'onoma
   de M\'exico}  \\
 {Apartado Postal 70-543, Cuidad de M\'exico 04510, Mexico }
}

\begin{document}

\maketitle
\begin{abstract}
  We investigate whether the Balitsky-Fadin-Kuraev-Lipatov (BFKL) low
  $x$ evolution equation is capable to describe the energy dependence
  of the exclusive photo-production cross-section of vector mesons
  $J/\Psi$ and $\Upsilon$ on protons. Such cross-sections have been
  measured by both HERA experiments H1 and ZEUS in electron-proton
  collisions and by LHC experiments ALICE, CMS and LHCb in
  ultra-peripheral proton-proton and ultra-peripheral proton-lead
  collisions. Our approach provides a perturbative description of the
  rise with energy and relies only on a fit of the initial transverse
  momentum profile of the proton impact factor, which can be extracted
  from BFKL fits to inclusive HERA data. We find that BFKL evolution
  is capable to provide a very good description of the energy
  dependence of the current data set, while the available fits of the
  proton impact factor require an adjustment in the overall
  normalization.
\end{abstract}

\section{Introduction}
\label{sec:introduction}

The Large Hadron Collider (LHC) provides due to its large center of
mass energy a unique opportunity to explore the high energy limit of
Quantum Chromodynamics (QCD).  In the presence of a hard scale the
theoretical description of the latter is based on the
Balitsky-Fadin-Kuraev-Lipatov (BFKL) evolution equation \cite{BFKL1},
currently known up to next-to-leading order (NLO) \cite{Fadin:1998py}
in the strong coupling constant $\alpha_s$.  The bulk of searches for
BFKL dynamics at the LHC concentrates  on the analysis of
correlations in azimuthal angels of jets, with the most prominent
example the angular decorrelation of a pair of forward-backward or
Mueller-Navelet jets.  Data collected for such angular correlations
during the $7$ TeV run provide currently first
phenomenological evidence for BFKL dynamics at the LHC
\cite{Ducloue:2013bva}.  More recent attempts on the theory side
include now also the study of angular correlations of up to 4 jets,
which are expected to provide further inside into  BFKL dynamics
and the realization of  so-called Multi-Regge-Kinematics at the LHC
\cite{Caporale:2015vya}.\\

The study of angular decorrelation is very attractive from a theory
point of view,  since the resulting perturbative description is very
stable and only weakly affected by soft and collinear radiative
corrections. At the same time such angular decorrelations allow only
to probe components of the BFKL kernel associated with non-zero
conformal spin, $n \neq 0$. The  exploration of $n \neq 0$
components is of interest in
its own right and further  allows already to test the calculational
framework of high-energy factorization, which underlies the formulation
of BFKL evolution. On the other hand these studies do not allow to address one of the
central questions of the QCD high energy limit, namely the growth of
perturbative cross-sections with energy. The latter corresponds to the
so-called `hard' Pomeron and is governed by the the conformal spin
zero $n=0$ component of the BFKL kernel.    \\

Unlike the $n \neq 0$ terms, the $n=0$ component is strongly affected
by large (anti-) collinear logarithms which need to be
resummed. Building on \cite{Salam:1998tj, Vera:2005jt}, such a
resummed NLO BFKL kernel has been constructed in
\cite{Hentschinski:2012kr,Hentschinski:2013id}, and employed for a fit
to proton structure functions measured in inclusive Deep Inelastic
Scattering (DIS) at HERA. These results have then been subsequently
used to extract an unintegrated gluon density and to provide
predictions for rapidity and transverse momentum distributions of
forward $b$-jets at the LHC in \cite{Chachamis:2015ona}. In the
following we will study the cross-section for exclusive
photo-production of vector mesons $J/\Psi$ and $\Upsilon$ on a
proton. In particular we are interested on a description of the rise
with center-of-mass energy $W$ of the $\gamma p \to V p$ cross-section
($V = J/\Psi, \Upsilon$), combining measurements at HERA and the LHC.
At the LHC the cross-sections for the process $\gamma p \to V p$ can
be extracted from ultra-peripheral proton-proton ($pp$) and
proton-lead ($pPb$) collisions, which allow to test the gluon
distribution in the proton down to very small values of the proton
momentum fraction $x > 4 \cdot 10^{-6}$.  For this processes the mass
of the heavy quarks, {\it i.e. } charm ($J/\Psi$) and bottom
($\Upsilon$), provide the hard mass-scale which allows for a
description within perturbative QCD. \\

Currently there exist various studies of the data collected during the
$7/8$ TeV run
\cite{Jones:2013eda,Jones:2013pga,Goncalves:2015pki,Fiore:2014oha,Armesto:2014sma,Goncalves:2014wna,Goncalves:2014swa,Schafer:2007mm},
see also \cite{Costa:2012fw,Goncalves:2009gs}.  In following we focus
on the question whether perturbative BFKL evolution is capable to
describe the rise of the $\gamma p \to V p$ cross-section with energy
{\it i.e.} we investigate whether the observed rise can be described
purely perturbatively, avoiding both the use of a fitted $W$
dependence as well as ideas related to gluon saturation.  While our
description involves necessarily also a fit of initial conditions to
data, this fit is restricted to the transverse momentum distribution
inside the proton, which has been determined in the analysis of HERA
data in \cite{Hentschinski:2012kr,Hentschinski:2013id}. The
$W$-dependence arises on the other hand due to a solution of the NLO
BFKL equation with collinear improvements, combined with an optimal
renormalization
scale setting.\\

The outline of this paper is as follows: In
Sec.~\ref{sec:vect-meson-prod} we present the theoretical framework of
our study, including a short review of the unintegrated BFKL gluon
density of \cite{Chachamis:2015ona} and a determination of the vector
meson photo-production impact factor. In
Sec.~\ref{sec:numerical-results} we present the numerical results of
our study, including a comparison to data while we present in
Sec.~\ref{sec:disc-concl} our conclusions and an outlook on future
work. Two integrals needed for the derivation of the impact factor are
collected in the Appendix.

\section{Vector meson production in the high energy limit}
\label{sec:vect-meson-prod}

In the following we describe the framework on which our study is based
on. We study the process
\begin{align}
  \label{eq:30}
 \gamma(q) + p(p) & \to V(q') + p(p') \, ,
\end{align}
where $V = J/\Psi, \Upsilon(1S)$ while $\gamma$ denotes a quasi-real
photon with virtuality $Q \to 0$; $W^2 = (q + p)^2$ is the squared
center-of-mass energy of the $\gamma(q) + p(p)$
collision. Neglecting proton mass effects, {\it i.e.} working in the
limit $q^2 = 0 = p^2$, the following Sudakov decomposition holds for
the final state momenta in the high energy limit $W \gg M_V$,
\begin{align}
  \label{eq:31}
q' & = q + \frac{M_V^2 + {\bm \Delta}^2}{W^2} p + { \Delta}_t &
p' & = p + \frac{{\bm \Delta}^2}{W^2} q  -  { \Delta}_t
\end{align}
with $l_t^2 = - {\bm l}^2$ and $l_t \cdot p = 0 = l_t \cdot q$ for a
generic momentum $l$. With the momentum transfer
$t = (q-q')^2 = -{\bm \Delta}^2$, the differential cross-section for
the exclusive production of a vector meson can be written in the
following form
\begin{align}
  \label{eq:16}
  \frac{d \sigma}{d t} \left(\gamma p \to V p \right)
& =
\frac{1}{16 \pi} \left|\mathcal{A}_{T, L}^{\gamma p \to V p}(W^2, t) \right|^2 \, .
\end{align}
where $\mathcal{A}(W^2, t)$ denotes the scattering amplitude for the
reaction $\gamma p \to V p$ for color singlet exchange in the
$t$-channel, with an overall factor $W^2$ already extracted.
\begin{figure}[t]
  \centering
  \includegraphics[width=.4\textwidth]{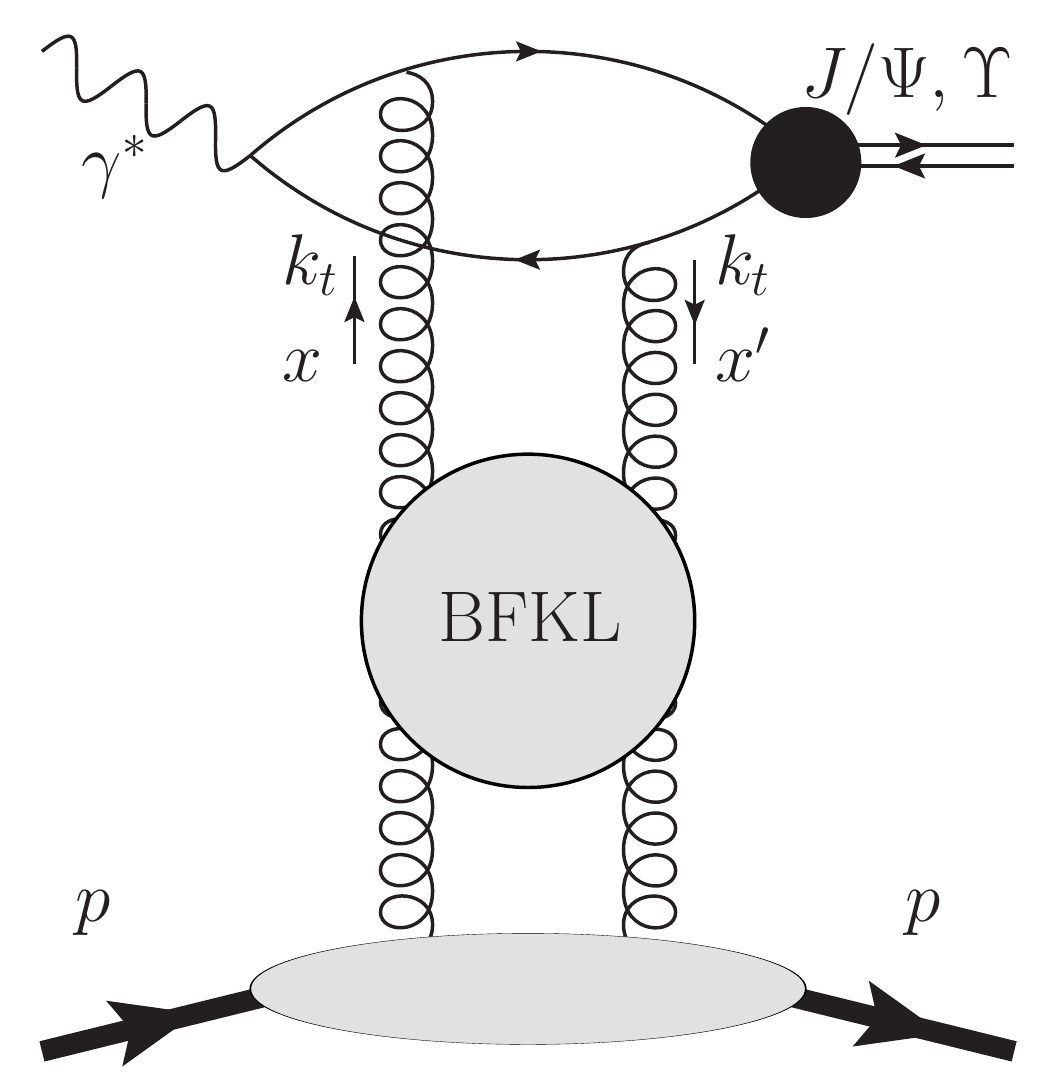}
  \caption{\it Schematic picture of the high energy factorized
    amplitude for photo-production of vector mesons $J/\Psi, \Upsilon$
  with zero momentum transfer $t$ = 0. In the high energy limit the amplitude factorizes into the  impact factor for the transition $\gamma^* \to J/\Psi, \Upsilon $ (quark-loop), the BFKL-Green's function (central blob) and non-perturbative proton impact factor (lower blob).}
  \label{fig:amplitude}
\end{figure}
Within high energy factorization {\it i.e.} discarding terms
$\sim M_V^2/W^2$, this scattering amplitude can be written as a
convolution in transverse momentum space of the universal BFKL
Green's-function, which achieves a resummation of high energy
logarithms $\ln W^2$ to all orders in the strong coupling constant
$\alpha_s$, and two process-dependent impact factors which describe
the coupling of the Green's function to external states, see
Fig.~\ref{fig:amplitude}.  In the present case, one of the impact
factors describes the transition $\gamma \to V$ and is characterized
by the heavy quark mass $m_c$ and $m_b$ respectively, which provide
the hard scale of the process. The second impact factor, which describes
the transition  $p \to p$ is of non-perturbative origin; it
 needs to be modeled with free parameters to be fixed by a
fit to data. \\

In the high energy limit $W^2 \gg M_V^2$, this scattering amplitude is
dominated by its imaginary part,
$\mathcal{A}(W^2, t) \simeq i\cdot \Im\text{m} \mathcal{A}(W^2, t)$,
with the real part suppressed by powers of $\alpha_s$. Limiting
ourselves for the moment to the dominant imaginary part we find that
for the case of zero momentum transfer, $t = -{\bm \Delta}^2 = 0$, the
non-perturbative proton impact factor coincides for this process with
the corresponding proton impact factor found in fits to Deep-Inelastic
Scattering data. Such a fit of the forward $t=0$ proton impact factor
has been performed in \cite{Hentschinski:2012kr, Hentschinski:2013id}
which can be therefore used for phenomenological studies of vector
meson production.

\subsection{The NLO collinear improved BFKL unintegrated gluon density}
\label{sec:bfkl_gluon}

In \cite{Hentschinski:2012kr, Hentschinski:2013id} the following model
has been used for the proton impact factor
\begin{align}
  \label{eq:proton}
  \Phi_p\left(\frac{{\bm q}^2}{Q_0^2}, \delta \right) & =
\frac{\mathcal{C}}{  \pi \Gamma(\delta)}
 \left(\frac{{\bm q}^2}{Q_0^2} \right)^\delta e^{- \frac{{\bm q}^2}{Q_0^2}}\, .
\end{align}
The model introduces 2 free parameters plus an overall normalization
factor and provides a Poisson-distribution peaked at
${\bm q}^2 = \delta Q_0^2$.
\begin{table}[tb]
  \centering
  \begin{tabular}[tb]{c|c|c|c|c}
\hline \hline
  &  virt. photon impact factor & $Q_0$/GeV & $\delta$ & $\mathcal{C}$ \\
\hline
  fit 1 &   leading order (LO)  & $ 0.28$  &  $ 8.4$ & $  1.50$ \\
\hline
fit 2 & LO with kinematic improvements &   $ 0.28$  &  $ 6.5$ & $  2.35 $  \\
  \end{tabular}
  \caption{\it Parameters of the proton impact factor obtained in
    \cite{Hentschinski:2013id} through a fit to combined HERA data }
  \label{tab:fit}
\end{table}
Depending on the precise form of the virtual photon impact factor, two
sets of parameters have been determined, which are summarized in
Tab.~\ref{tab:fit}, where for the second fit the leading order virtual
photon impact factor has been supplemented with DGLAP inspired
kinematic corrections \cite{Kwiecinski:1997ee}; both
fits have been performed for $n_f = 4$ mass-less flavors.\\

In \cite{Chachamis:2015ona} the results of this fit have been used to
introduce a NLL BFKL unintegrated gluon density as the following
convolution of proton impact factor and BFKL Green's function
\begin{align}
  \label{eq:bfkl_gluon_schematic}
  G(x, {\bm k}^2, Q_0^2) & = \int \frac{d {\bm q}^2}{{\bm q}^2} \mathcal{F}^{\text{DIS}}(x, {\bm k}^2, {\bm q}^2)  \Phi_p\left( \frac{{\bm q}^2}{Q_0^2}\right) \, .
\end{align}
In  Mellin space conjugate to transverse momentum space this unintegrated gluon density can be written as
\begin{align}\label{eq:Gudg}
  G\left(x, {\bm k}^2, M\right)
& =
 \frac{1}{{\bm k}^2}\int \limits_{\frac{1}{2}-i\infty}^{\frac{1}{2} + i \infty}  \frac{d \gamma}{2 \pi i}   \; \; \hat{g}\left(x, \frac{M^2}{Q_0^2}, \frac{\overline{M}^2}{M^2}, \gamma \right) \, \, \left(\frac{{\bm k}^2}{Q_0^2} \right)^\gamma
\end{align}
where $M$ is a characteristic hard scale of the process which in the
case of the DIS fit has been identified with the virtuality of the
photon and $\overline{M}$ is a corresponding scale which enters the
running coupling constant (see also the discussion below); in the DIS
analysis $ M = \overline{M}$ and both scales have been identified with
the virtuality of the scattering photon.  $\hat{g}$ is finally an
operator in $\gamma$ space and defined as
\begin{align}
  \label{eq:23}
 \hat{g}\left(x, \frac{M^2}{Q_0^2},  \frac{\overline{M}^2}{M^2},  \gamma \right)
  & =
 \frac{\mathcal{C}\cdot \Gamma(\delta - \gamma)} {\pi \Gamma(\delta)}  \; \cdot \;
 \left(\frac{1}{x}\right)^{\chi\left(\gamma,  \frac{\overline{M}^2}{M^2} \right)} \,\cdot \notag
 \\
&
  \Bigg\{1    + \frac{\bar{\alpha}_s^2\beta_0  \chi_0 \left(\gamma\right) }{8 N_c} \log{\left(\frac{1}{x}\right)}
  \Bigg[- \psi \left(\delta-\gamma\right)
 +  \log \frac{{M}^2}{Q_0^2} -  \partial_\gamma \Bigg]\Bigg\}\;,
\end{align}
where $\bar{\alpha}_s = \alpha_s N_c/\pi$ with $N_c$ the number of
colors and $\chi(\gamma, \overline{M}^2/M^2)$ is the next-to-leading
logarithmic (NLL) BFKL kernel after collinear improvements; in
addition large terms proportional to the first coefficient of the QCD
beta function, $\beta_0 =  11 N_c/3 - 2 n_f /3$ have been resumed
through employing a Brodsky-Lepage-Mackenzie (BLM) optimal scale
setting scheme ~\cite{Brodsky:1982gc}.  The NLL kernel with collinear
improvements reads
 \begin{align}\label{eq:gluongf}
\chi\left(\gamma,  \frac{\overline{M}^2}{M^2}\right) &=
{\bar\alpha}_s\chi_0\left(\gamma\right)+
{\bar\alpha}_s^2\tilde{\chi}_1\left(\gamma\right)-\frac{1}{2}{\bar\alpha}_s^2
\chi_0^{\prime}\left(\gamma\right)\chi_0\left(\gamma\right)
\notag \\
&\hspace{2cm} +
\chi_{RG}({\bar\alpha}_s,\gamma,\tilde{a},\tilde{b}) - \frac{\bar{\alpha}_s^2 \beta_0}{8 N_c} \chi_0(\gamma)\log  \frac{\overline{M}^2}{M^2} .
\end{align}
with the leading-order BFKL eigenvalue,
\begin{align}
  \label{eq:chi0}
  \chi_0(\gamma) & = 2 \psi(1) - \psi(\gamma) - \psi(1-\gamma) \, .
\end{align}
We note that the last term in the second line of
Eq.~\eqref{eq:gluongf} was not present in the final results of
\cite{Hentschinski:2012kr,Hentschinski:2013id} and
\cite{Chachamis:2015ona}, but can be easily derived from an
intermediate result provided in \cite{Hentschinski:2012kr}. It has
been re-introduced to assess possible uncertainties of the final
result due to identifying $\overline{M} = M$. The term responsible for
the resummation of collinear enhanced terms reads
\begin{align}
\label{eq:RG}
 \chi_{RG}(\bar{\alpha}_s, \gamma, a, b)
&  =  \,\,\bar{\alpha}_s (1+ a \bar{\alpha}_s) \left(\psi(\gamma)
- \psi (\gamma-b \bar{\alpha}_s)\right)  - \frac{\bar{\alpha}_s^2}{2}
  \psi'' (1-\gamma)  -  \frac{ b \bar{\alpha}_s^2 \cdot \pi^2}{\sin^2{(\pi
  \gamma)}}
\notag \\
&
+ \frac{1}{2} \sum_{m=0}^\infty \Bigg(\gamma-1-m+b \bar{\alpha}_s
  - \frac{2 \bar{\alpha}_s (1+a \bar{\alpha}_s)}{1-\gamma+m}
\notag \\
& \hspace{4cm}
+ \sqrt{(\gamma-1-m+b \bar{\alpha}_s)^2+ 4 \bar{\alpha}_s (1+a \bar{\alpha}_s)} \Bigg) \, .
\end{align}
For details on the derivation of this term we refer to the discussion
in   \cite{Hentschinski:2012kr}, see also \cite{Salam:1998tj, Vera:2005jt}.
Employing BLM optimal scale setting and the momentum space (MOM)
physical renormalization scheme based on a symmetric triple gluon
vertex~\cite{Celmaster:1979km} with $Y \simeq 2.343907$ and gauge
parameter $\xi =3$ one obtains the following next-to-leading order
BFKL eigenvalue
\begin{eqnarray}
\label{eq:chi1NLO}
\tilde{\chi}_1 (\gamma) &=& \tilde{\cal S} \chi_0 (\gamma) + \frac{3}{2} \zeta(3)
+ \frac{ \Psi ''(\gamma) + \Psi''(1-\gamma)- \phi(\gamma)-\phi (1-\gamma) }{4} \nonumber \\
&-& \frac{\pi^2 \cos{(\pi \gamma)}}{4 \sin^2{(\pi \gamma)}(1-2\gamma)}
\left[3+\left(1+\frac{n_f}{N_c^3}\right) \frac{2+3\gamma(1-\gamma)}{(3-2\gamma)(1+2\gamma)}\right] \nonumber\\
&+&\frac{1}{8} \left[\frac{3}{2} (Y-1) \xi
   +\left(1-\frac{Y}{3}\right) \xi ^2+\frac{17 Y}{2}-\frac{\xi ^3}{6}\right] \chi_0 (\gamma),
\end{eqnarray}
where $\tilde{\cal S} =\tfrac{ (4-\pi^2)}{12}$, see also the
discussion in \cite{Brodsky:2002ka}.  The coefficients
$\tilde{a}, \tilde{b}$ which enter the collinear resummation term
Eq.~\eqref{eq:RG} are obtained as the coefficients of the $1/\gamma$
and $1/\gamma^2$ poles of the NLO eigenvalue. In the case of
Eq.~\eqref{eq:chi1NLO} one has
\begin{align}
  \label{eq:aundb}
  \tilde{a} &=  - \frac{13}{36} \frac{n_f}{N_c^3}- \frac{55}{36} + \frac{3 Y - 3}{16}\xi + \frac{3 - Y}{24} \xi^2 - \frac{1}{48}\xi^3 + \frac{17}{16}Y \notag\\
\tilde{b} &= - \frac{n_f}{6N_c^3}- \frac{11}{12}.
\end{align}
Employing   BLM optimal scale setting, the running coupling constant becomes  dependent on the Mellin-variable $\gamma$ and reads
\begin{eqnarray}
\label{eq:BLM}
\tilde{\alpha}_s \left( \overline{M} \cdot Q_0, \gamma \right) &=&
                                                                   \frac{4 N_c}{\beta_0 \left[\log{\left(\frac{\overline{M}\cdot  Q_0}{ \Lambda^2}\right)}
+\frac{1}{2} \chi_0 (\gamma) - \frac{5}{3} +2 \left(1+ \frac{2}{3} Y\right)\right]},
\end{eqnarray}
In addition, in order to access the region of small photon virtualities, in \cite{Hentschinski:2012kr,Hentschinski:2013id}, a parametrization of the running coupling introduced by Webber in
Ref.~\cite{Webber:1998um} has been used,
 \begin{eqnarray}
\alpha_s \left(\mu^2\right) =  \frac{4\pi}{\beta_0\ln{\frac{\mu^2}{\Lambda^2}}}
+ f\left(\frac{\mu^2}{\Lambda^2}\right) , \;\;\;\; f\left(\frac{\mu^2}{\Lambda^2}\right) =  \frac{4\pi}{\beta_0}\; \frac{ 125\left(1 + 4 \frac{\mu^2}{\Lambda^2}\right)}{\left(1 - \frac{\mu^2}{\Lambda^2}\right)\left(4 + \frac{\mu^2}{\Lambda^2}\right)^4},
\end{eqnarray}
with $\Lambda=0.21\;$GeV. At low scales this modified running coupling
is consistent with global data of power corrections to perturbative
observables, while for larger values it coincides with
the conventional perturbative running coupling constant. For further details we refer the interested reader to  \cite{Hentschinski:2012kr,Hentschinski:2013id} and references therein.  \\

\subsection{The vector meson photo-production impact factor}
\label{sec:impactVM}

To use the above unintegrated gluon density for the description of the
process $\gamma p \to V p $, we still require the impact factor for
the transition $\gamma \to V$. To the best of our knowledge, such an
impact factor is currently not known within the BFKL framework. It is
however possible to extract the required quantity from a description
based on a factorization of the amplitude in the high energy limit
into light-front wave function and and dipole amplitude. In the dilute
limit, the factorization into light-front wave function and dipole
amplitude becomes equivalent to the factorization into impact factor
and unintegrated gluon density and it is therefore possible to recover
the required impact factor from these results.  Our starting point is
the following expression for the imaginary part of the vector meson
photo-production scattering amplitude \cite{Cox:2009ag,
  Kowalski:2006hc}
\begin{align}
\label{am-i}
\Im \text{m}\mathcal{A}^{\gamma^* p\rightarrow
    Vp}_{T,L} (W, t) &= 2 \, \int\!d^2{\bm r}\int\! d^2{\bm b}
  \int_0^1\! \frac{d {z}}{4 \pi }\;(\Psi_{V}^{*}\Psi)_{T,L}
  \;{e}^{-i [\bm{b}-(1-z)\bm{r}]\cdot\bm{\Delta}}\mathcal{N}\left(x,r,b\right),
\end{align}
where $\mathcal{N}\left(x,r,b\right)$ is the dipole amplitude and
$T, L$ denotes transverse and longitudinal polarization of the virtual
photon respectively and $t = - {\bm \Delta}^2$.  The overlap between
the photon and the vector meson light-front wave function reads
\begin{align}
  \label{eq:21}
  \left(\Psi_{V}^{*}\Psi\right)_{T}(r) & = \hat{e}_f e \frac{N_c}{\pi z (1-z)}
 \bigg\{
 m_f^2 K_0(\epsilon r) \phi_T(r,z) - \left[z^2 + (1-z)^2 \right] \epsilon K_1(\epsilon r) \partial_r \phi_T(r,z)
\bigg\}
\notag \\
 \left(\Psi_{V}^{*}\Psi\right)_{L} (r)& = \hat{e}_f e \frac{N_c}{\pi} 2 Q z(1-z) K_0(\epsilon r)
\bigg\{
M_V \phi_L(r,z) + \delta  \frac{m_f^2 - \nabla^2_r}{M_v z (1-z)}\phi_L(r,z)
\bigg\} \, ,
\end{align}
where from now on we discard longitudinal photon polarizations since
the corresponding wave function overlap is vanishing in the limit
$Q\to 0$ in which we are working.  To keep our result applicable to
the case $Q \neq 0$, we however keep on using the notation
$\epsilon^2 \equiv z(1-z)Q^2+m_f^2$, with $\epsilon^2 = m_f^2$ for
real photons. Furthermore $r = \sqrt{{\bm r}^2}$, while $f = c, b$
denotes the flavor of the heavy quark, with charge  $\hat{e}_f=2/3$, $1/3$,
corresponding to $J/\psi$ and $\Upsilon$ mesons respectively.  For the
scalar parts of the wave functions $\phi_{T,L}(r,z)$, we follow
closely \cite{Armesto:2014sma} and employ the boosted Gaussian
wave-functions with the Brodsky-Huang-Lepage prescription
\cite{Brodsky:1980vj}.  For the ground state vector meson ($1s$) the
scalar function $\phi_{T,L}(r,z)$, has the following general form
\cite{Cox:2009ag, Nemchik:1994fp},
\begin{align}
\label{eq:1s_groundstate}
\phi_{T,L}^{1s}(r,z) &= \mathcal{N}_{T,L} z(1-z)
  \exp\left(-\frac{m_f^2 \mathcal{R}_{1s}^2}{8z(1-z)} - \frac{2z(1-z)r^2}{\mathcal{R}_{1s}^2} + \frac{m_f^2\mathcal{R}_{1s}^2}{2}\right)  \, .
\end{align}
The free parameters $N_T$ and $\mathcal{R}_{1s}$ of this model have been
determined in various studies from the normalization condition of the
wave function and the decay width of the vector mesons. In the
following we use the most recent available values {\it i.e.}
\cite{Armesto:2014sma} (for the $J/\Psi$) and \cite{Goncalves:2014swa}
(for the $\Upsilon$) The results are summarized in Tab.~\ref{vm_fit}.
\begin{table}
\centering
\begin{tabular}{c|c|c|c|c|c|c}
\hline\hline &&&&&& \vspace{-.2cm}\\
Meson & $m_f/\text{GeV}$  & $\mathcal{N}_T$ &
                                              $\mathcal{R}^2$/$\text{GeV}^{-2}$
  & $M_V$/GeV  & $ 8 {\cal R^{-2}}/\text{GeV}^2$ &  $\frac{1}{4}
                                                      M_V^2/\text{GeV}^2$\\
&&&&&& \vspace{-.2cm}\\  \hline
$J/\psi$ & $m_c=1.27$&   $0.596$ & $2.45$ & $3.097$  & $3.27$ & $2.40$\\ \hline
$\Upsilon$ & $m_b = 4.2 $ & $0.481$ & $0.57$ & $9.460$ & $15.38$  & $22.42$\\ \hline

\end{tabular}
\caption{Parameters of the boosted Gaussian vector meson wave functions for $J/\psi$ and $\Upsilon$ obtained in  \cite{Armesto:2014sma, Goncalves:2014swa}. The last two  columns give the two possible  hard scales used in  the BFKL analysis.}
\label{vm_fit}
\end{table}
In the forward limit $t=0$, the entire dependence of the integrand on
the impact parameter ${\bm b}$ is contained in the dipole amplitude
which results into the following inclusive dipole cross-section,
\begin{align}
  \label{eq:intN}
  2 \int d^2 {\bm b} \,   \mathcal{N}\left(x,r,b\right) & = \sigma_0 N(x, {\bm r})\,.
\end{align}
The relation between the latter and an  unintegrated gluon density
 has been worked  in \cite{Kutak:2004ym}  and is given by
\begin{align}
  \label{eq:Nfromugd}
  \sigma_0 N({\bm r}, x) & = \frac{4 \pi}{N_c} \int \frac{d^2 {\bm
                           k}}{{\bm k}^2}
\left(1-e^{i {\bm k}\cdot {\bm r}}\right) \alpha_s  G(x, {\bm k}^2) \, .
\end{align}
 This expression can then be used to  calculate the BFKL impact factor from the light-front wave function overlap Eq.~\eqref{eq:21}. In particular we find
\begin{align}
 \label{dipole-to-bfkl}
 \Im\text{m} \mathcal{A}^{\gamma^* p\rightarrow Vp}_{T}  (W, 0)
&  = \,
\int\!d^2{\bm r}
\int_0^1\! \frac{d {z}}{4 \pi }\;(\Psi_{V}^{*}\Psi)_{T} (r)\;  \cdot   \sigma_0 {N}\left(x, r \right)
\notag \\
&\hspace{-2.3cm} =    \alpha_s(\overline{M}\cdot  Q_0)  \int\limits_{\frac{1}{2} - i \infty}^{\frac{1}{2} + i \infty} \frac{d \gamma}{2 \pi i } \int\limits_0^1 \frac{d {z}}{4 \pi }  \;
 \;\hat{g}\left(x, \frac{M^2}{Q_0^2}, \frac{\overline{M}^2}{M^2}, Q_0,
  \gamma\right)  \cdot    \Phi_{V,T}(\gamma, z, M ) \cdot  \left(  \frac{M^2}{Q_0^2}\right)^\gamma.
\end{align}
In the above expression, $M$ and $\overline{M}$  are the mass-scales
introduced in Eq.~\eqref{eq:23}. The scale of the strong coupling
$\alpha_s$ in Eq.~\eqref{eq:Nfromugd}, \eqref{dipole-to-bfkl} has been
set in accordance with the conventions used in the HERA fit\footnote{A
  precise determination of the scale of this  running coupling would
  require the complete NLO corrections to the impact factor which are
  currently not available} \cite{Hentschinski:2013id}.   From
Eq.~\eqref{dipole-to-bfkl} we obtain
\begin{align}
  \label{eq:24}
   \Phi_{V,T}&(\gamma, z, M)  =
\frac{4 \pi}{N_c} \int d^2 {\bm r} \int \frac{d^2 {\bm k}}{\left({\bm
               k}^2\right)^2} \left(1 - e^{i {\bm k}\cdot {\bm r}}
               \right)
\left(\frac{{\bm k}^2}{M^2} \right)^\gamma \left(\Psi_V^* \Psi \right)_T(r)
\notag \\
&
= e \hat{e}_f 8 \pi^2  \mathcal{N}_T \frac{\Gamma(\gamma) \Gamma(1-\gamma)}{m_f^2} \left(\frac{m_f^2 \mathcal{R}^2}{8 z(1-z)} \right)^2
e^{- \frac{m_f^2 \mathcal{R}^2}{8 z(1-z)} }
e^{\frac{m_f \mathcal{R}^2}{2}}
\left(\frac{8z(1-z)}{M^2 \mathcal{R}^2} \right)^\gamma \notag \\
&
\bigg[
U\left(2-\gamma, 1, \frac{\epsilon^2 \mathcal{R}^2}{8z(1-z)}\right)
+
 [z^2 +
  (1-z)^2] \frac{\epsilon^2 (2-\gamma)}{2 \cdot m_f^2}
U\left(3-\gamma, 2, \frac{\epsilon^2 \mathcal{R}^2}{8z(1-z)}\right)
\bigg]\, ,
\end{align}
where $U(a,b,z)$ is a hypergeometric function of the second kind or
Kummer's function. Some useful integrals in the derivation of this
result are summarized in the appendix. Expanding
Eq.~\eqref{dipole-to-bfkl} to NLO in $\alpha_s$, it is straightforward
to verify that our result is independent of $M$ to NLO
accuracy. Furthermore one can verify that the resummed BFKL eigenvalue
Eq.~\eqref{eq:gluongf} is furthermore independent of the choice of
$\overline{M}$ up to terms $\mathcal{O}(\alpha_s^3)$.

\subsection{Real part, phenomenological corrections and integrated cross-sections}
\label{sec:scatt-ampl-high}

Even though the real part of the scattering amplitude is suppressed by
powers of $\alpha_s$ in the high energy limit, it can still provide a
sizable correction to the cross-section and should be therefore
included. In the high energy limit it is possible to obtain this real
part from the imaginary part  using dispersion relation. One has
\begin{align}
  \label{eq:32}
  \frac{\Re\text{e} \mathcal{A}(W^2, t)}{\Im\text{m} \mathcal{A}(W^2, t)}
&=
\tan \frac{\lambda \pi }{2},
& \text{with}&
& \lambda & = \frac{d \ln  \mathcal{A}(W^2, t) }{ d \ln W^2} \, .
\end{align}
Eq.~\eqref{eq:32} is  frequently used in the literature in the study
of photo-production of vector mesons. Within our framework we
write first the imaginary part of the scattering amplitude  as a double Mellin transform
\begin{align}
  \label{eq:doubleMellin_IMofA}
 \Im\text{m}& \mathcal{A}_T^{\gamma p \to V p} (W^2, 0)
  =
 \alpha_s(\overline{M}\cdot Q_0) \cdot    \int\limits_{\delta -
    i\infty}^{\delta + i \infty} \frac{d \omega}{2 \pi i}  \left(\frac{1}{x}\right)^\omega   \int\limits_{\frac{1}{2} - i\infty}^{\frac{1}{2} + i \infty} \frac{d \gamma}{ 2 \pi i}
\left(\frac{M^2}{Q_0^2} \right)^\gamma
    \notag \\
&
\int\limits_0^1 \frac{d z}{ 4 \pi} \,  \Phi_{V,T}(\gamma, z, M)
 \frac{\mathcal{C}\cdot \Gamma(\delta - \gamma)} {\pi \Gamma(\delta)}  \; \cdot \;
\,
  \Bigg\{\frac{1}{\omega - {\chi\left(\gamma, \frac{\overline{M}^2}{M^2} \right)} }
\notag \\
& \hspace{3cm} +  \frac{ {\bar{\alpha}_s^2\beta_0  \chi_0 \left(\gamma\right) }/({8 N_c}) }{ \left[  \omega - {\chi\left(\gamma, \frac{\overline{M}^2}{M^2} \right)}  \right]^2    }
  \Bigg[- \psi \left(\delta-\gamma\right)
 -\frac{d \ln \left[  \Phi_{V,T}(\gamma, z, M) \right] }{d \gamma }   \Bigg]\Bigg\}\, ,
\end{align}
\begin{figure}[t]
  \centering
  \includegraphics[width=.45\textwidth]{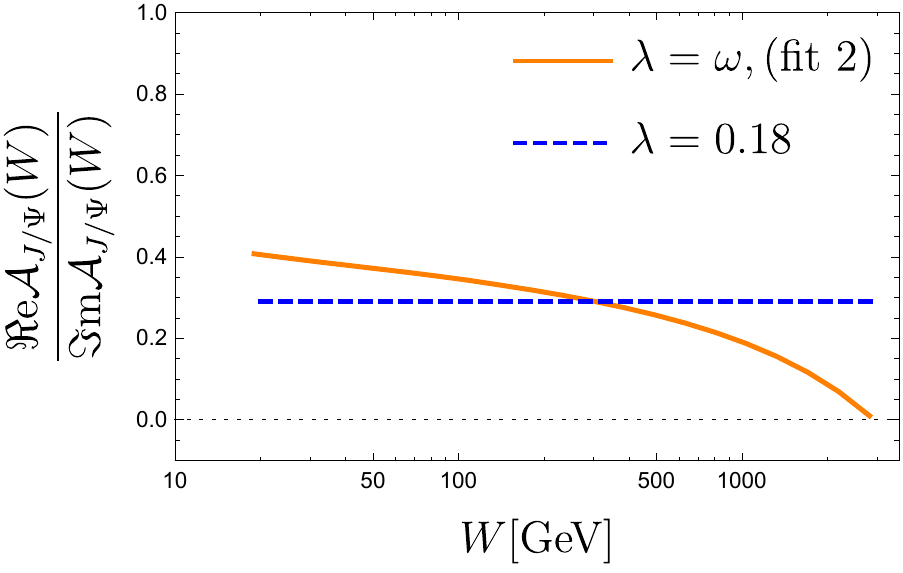}\hspace{1cm}
 \includegraphics[width=.45\textwidth]{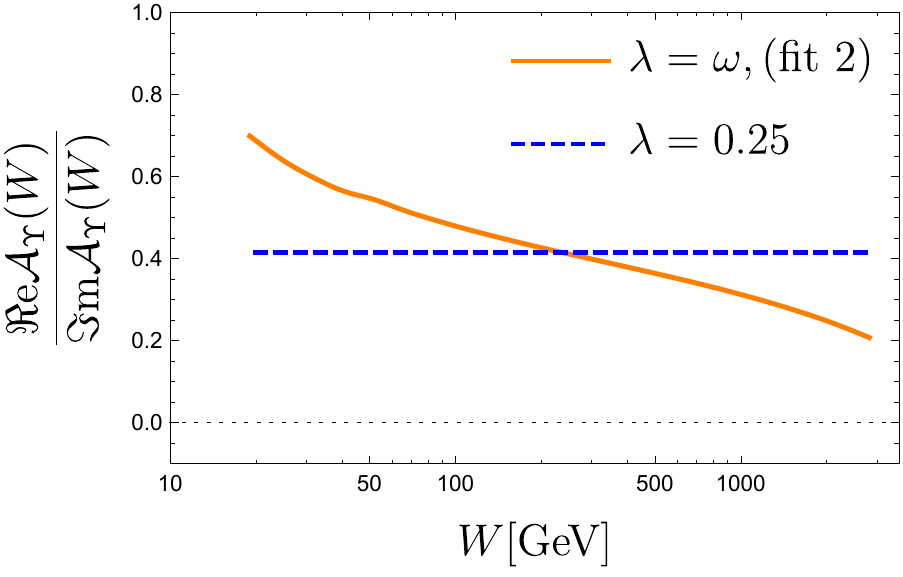}
  \caption{\it Ratio of real and imaginary part of the $\gamma p \to V p$ scattering amplitude for $V = J/\Psi$ (left) and $V = \Upsilon$ (right) as obtained from using a constant value (dashed) for $\lambda$ in Eq.~\eqref{eq:32} and the identification $\lambda = \omega$, see Eq.~\eqref{eq:doubleMellin_IMofA} and the subsequent discussion (continuous). }
  \label{fig:ratios}
\end{figure}
where the $\omega$-contour runs to the right of all singularities and
$x = \frac{M_V^2}{W^2 - m_p^2}$, with $m_p = .938$ GeV the proton
mass; note that $\mathcal{A}_L = 0$ due to $Q=0$. To determine the
real part we identify $\lambda$ with the Mellin variable conjugate to
the $x$, $\lambda = \omega$. The complete amplitude is then obtained
through multiplying the integrand of Eq.~\eqref{eq:doubleMellin_IMofA}
by a factor $ \left( i + \tan \frac{\omega \pi }{2} \right)$.  The
Mellin transform w.r.t. $\omega$ is then easily evaluated through
taking residues at the single and double pole at
$\omega = \chi(\gamma, \overline{M}^2.M^2)$, while residues at
$\omega = -1, -3, \ldots$ are subleading in the high energy/low $x$
limit and therefore neglected. As a consequence we obtain -- in
contrast to the bulk of phenomenological studies in the literature --
an energy dependent ratio of real and imaginary part, see
Fig.~\ref{fig:ratios} for numerical results. Particular for small
values of $W$, the real part provides a relative large correction, up
to $41\%$ in the case of the $J/\Psi$ and $70\%$ in the case of the
$\Upsilon$, see also the discussion in \cite{Baranov:2007zza}. On the
other hand, since this ratio is decreasing with increasing $W$, we
find that this energy-dependent ratio leads to a slow-down of the
growth with energy in the high energy region.
\\

Another phenomenological correction to the cross-section, which is
often included in studies of vector meson photo-production, arises due
to the fact that the proton momentum fractions $x$, $x'$ of the two
gluons coupling to the $\gamma \to V$ transition, can differ, even
though we are working in the forward limit $t=0$.  In
\cite{Shuvaev:1999ce} a corresponding corrective factor has been
determined for the case of the conventional (integrated) gluon
distribution, by relating the latter through a Shuvaev transform to
the generalized parton distribution (GPD).  Since we are dealing in
the current case with an transverse momentum dependent (unintegrated)
gluon density, such a corrective factor would at best be correct
approximately. Our numerical studies find no significant improvement
in the description of  data due to such a factor and we therefore
do not include it in our analysis. \\

While we calculated so far the differential cross-section at momentum
transfer $t=0$, experimental data which we wish to analyze, are
usually given for cross-sections integrated over $t$. It is therefore
necessary to model the $t$-dependence and to relate in this way the
differential cross-section at $t=0$ to the integrated
cross-section. Here we follow the prescription given in
\cite{Jones:2013eda,Jones:2013pga}, who assume an exponential drop-off
with $|t|$, $\sigma \sim \exp\left[-|t| B_D(W)\right]$ with an energy dependent
$t$ slope parameter $B_D$,  which can be  motivated by Regge theory,
\begin{align}
  \label{eq:18}
  B_D(W) & =\left[  b_0 + 4 \alpha' \ln \frac{W}{W_0} \right] \text{GeV}^{-2}.
\end{align}
For the numerical values we use $\alpha' = 0.06$ GeV$^{-2}$,
$W_0 = 90$ GeV and $b_0^{J/\Psi} = 4.9$ GeV$^{-2}$ in the case of the
$J/\Psi$, while $b_0^{\Upsilon} = 4.63$ GeV$^{-2}$ for $\Upsilon$
production, as proposed in \cite{Jones:2013eda,Jones:2013pga}.  The
total cross-section for vector meson production is therefore obtained
as
\begin{align}
  \label{eq:16total}
 \sigma^{\gamma p \to V p}(W^2) & = \frac{1}{B_D(W)} \frac{d \sigma}{d t} \left(\gamma p \to V p \right)\bigg|_{t=0}
.
\end{align}
For the sake of completeness we further provide our final expression for the differential cross-section at $t=0$. It is given by Eq.~\eqref{eq:16} for the case $t=0$ with
\begin{align}
  \label{eq:diffXsec_final}
  & \mathcal{A}_T^{\gamma p \to V p} (W^2, 0)
  =
 \alpha_s(\overline{M}\cdot Q_0) \cdot    \int\limits_{\frac{1}{2} - i\infty}^{\frac{1}{2} + i \infty} \frac{d \gamma}{ 2 \pi i}
\left(\frac{M^2}{Q_0^2} \right)^\gamma
    \left(i + \tan \frac{ \pi \cdot \chi(\gamma, \frac{\overline{M}^2}{M^2})}{2} \right) \notag \\
&
\int\limits_0^1 \frac{d z}{ 4 \pi} \Phi_{V,T}(\gamma, z, M)
 \frac{\mathcal{C}\cdot \Gamma(\delta - \gamma)} {\pi \Gamma(\delta)}  \; \cdot \;
 \left(\frac{1}{x}\right)^{\chi\left(\gamma,  \frac{\overline{M}^2}{M^2} \right)} \,\cdot
  \Bigg\{1    + \frac{\bar{\alpha}_s^2\beta_0  \chi_0 \left(\gamma\right) }{8 N_c}  \Bigg[ \ln{\left(\frac{1}{x}\right)}
\notag \\
&
 \qquad \qquad + \frac{\pi}{2} \left( \tan \left[\frac{ \pi \chi\left(\gamma, \frac{\overline{M}^2}{M^2}\right)}{2}  \right] - i\right)\Bigg]
  \Bigg[- \psi \left(\delta-\gamma\right)
 -\frac{d \ln \left[  \Phi_{V,T}(\gamma, z, M) \right] }{d \gamma }   \Bigg]\Bigg\}\, ,
\end{align}
where $x = \frac{M_V^2}{W^2 - m_p^2}$.

\section{Numerical results and Discussion}
\label{sec:numerical-results}

\begin{figure}[p!]
  \centering
   \includegraphics[width=.95\textwidth]{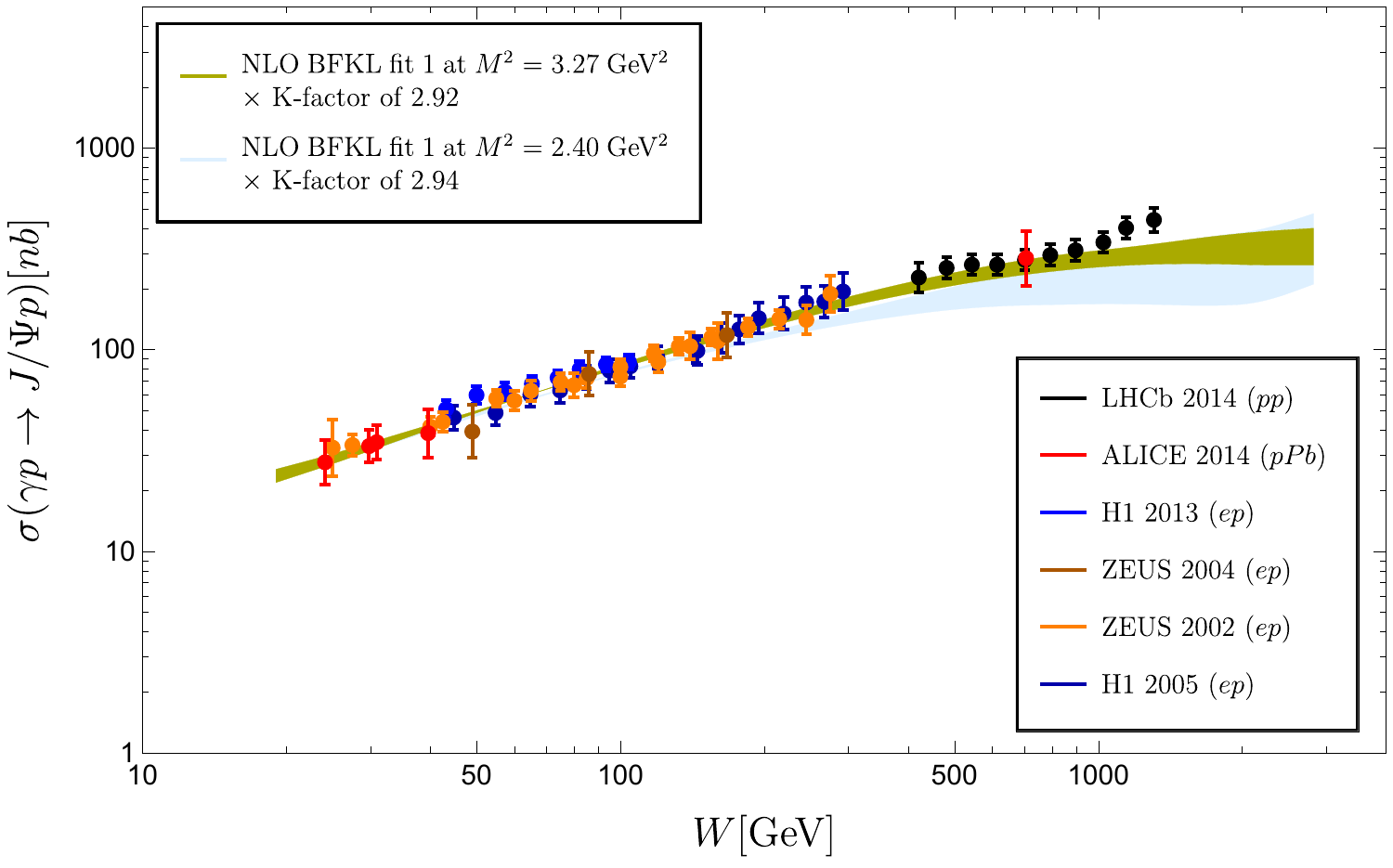}\\
\vspace{1cm}
  \includegraphics[width=.95\textwidth]{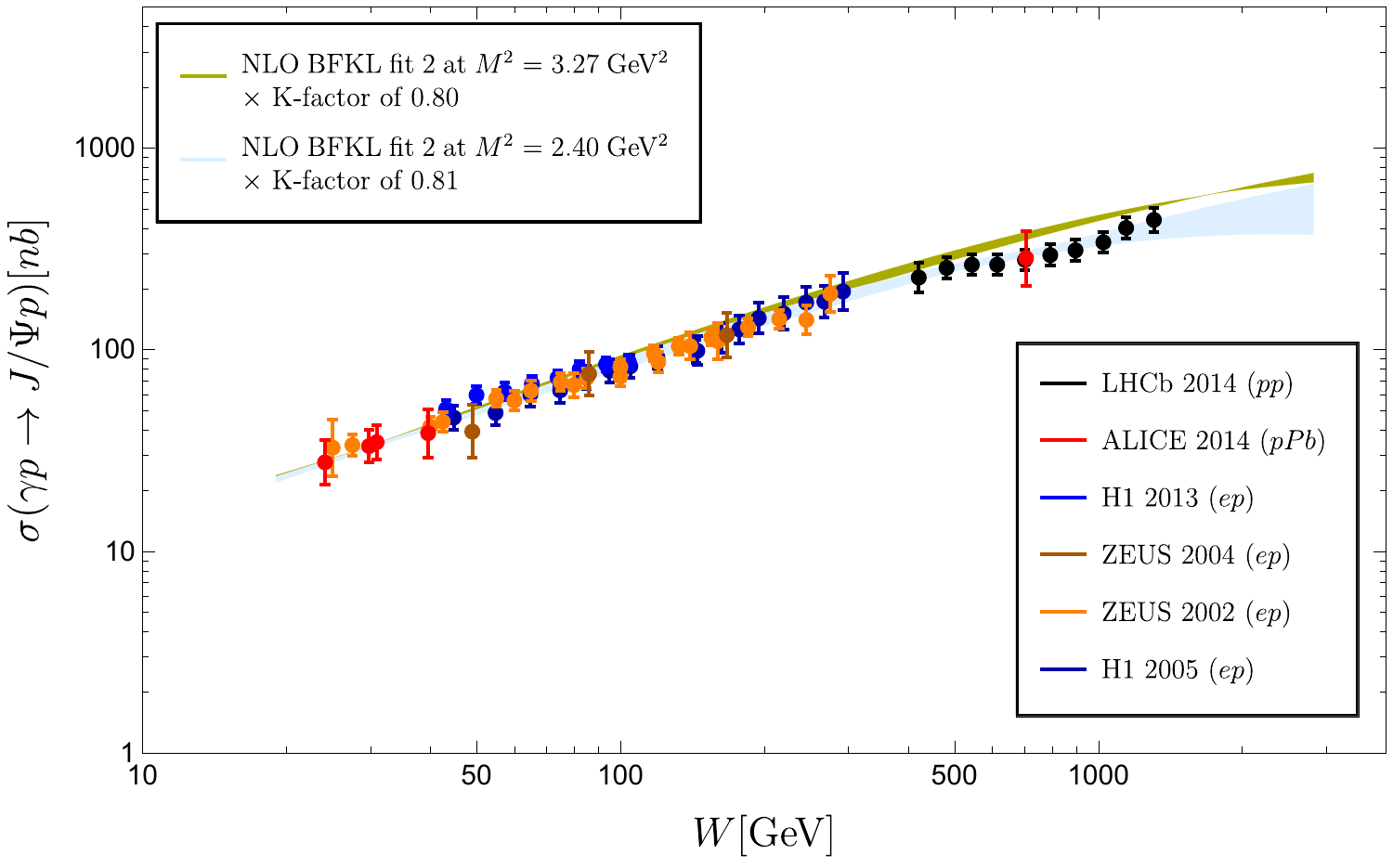}
  \caption{\it Energy dependence of the $J/\Psi$ photo-production
    cross-section as provided by the BFKL fit 1 (up) and 2 (down). The
    uncertainty band reflects a variation of the scale
    $\overline{M}^2 \to \{\overline{M}^2/2, \overline{M}^2 \cdot 2\}$.
    We also show photo-production data measured at HERA by ZEUS
    \cite{Chekanov:2002xi,Chekanov:2004mw} and H1
    \cite{Alexa:2013xxa,Aktas:2005xu} as well as LHC data obtained
    from ALICE \cite{TheALICE:2014dwa} and LHCb ($W^+$ solutions)
    \cite{Aaij:2013jxj}.}
  \label{fig:resultsJPsi}
\end{figure}
\begin{figure}[p!]
  \centering
  \includegraphics[width=.95\textwidth]{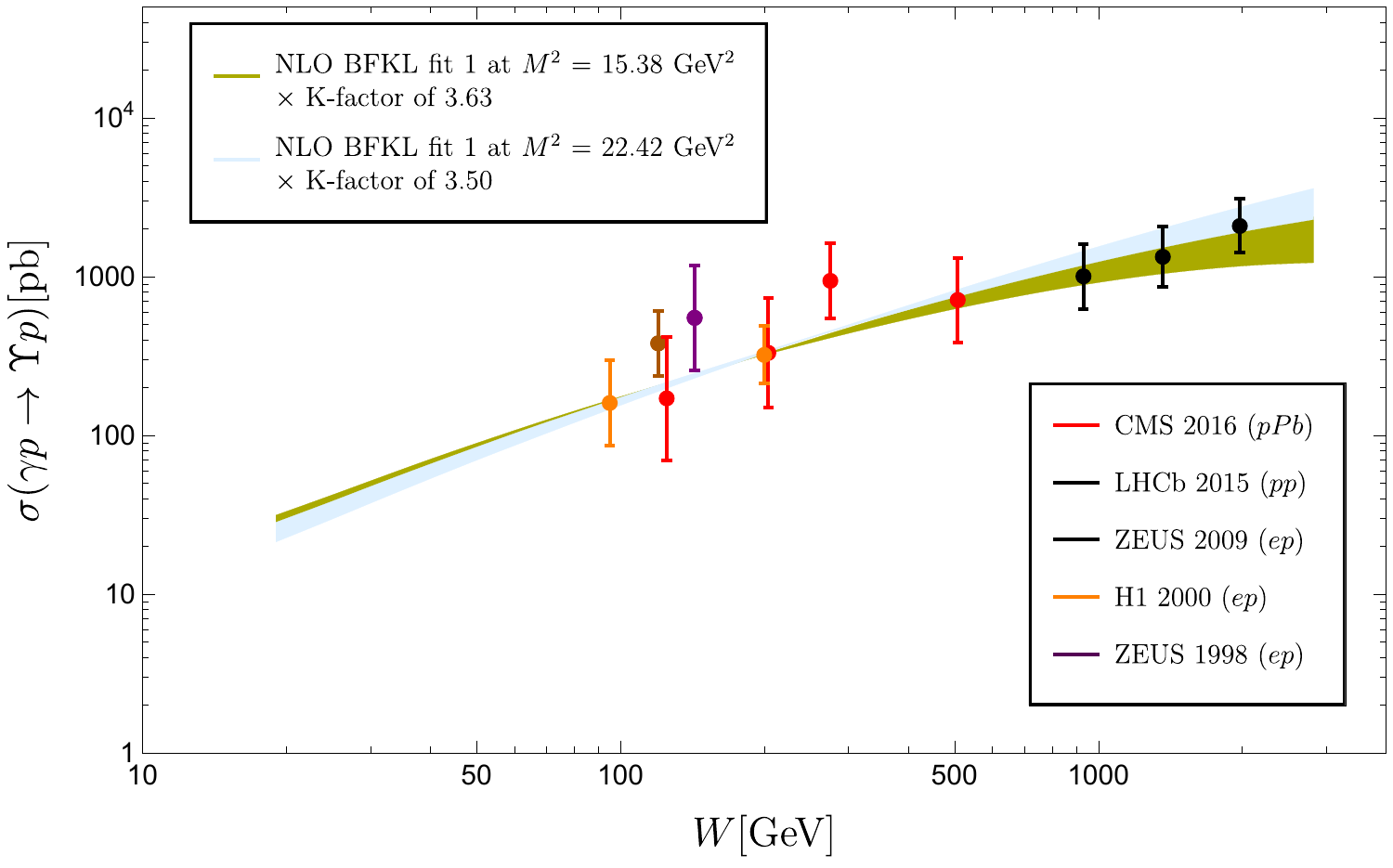}\\
\vspace{1cm}
  \includegraphics[width=.95\textwidth]{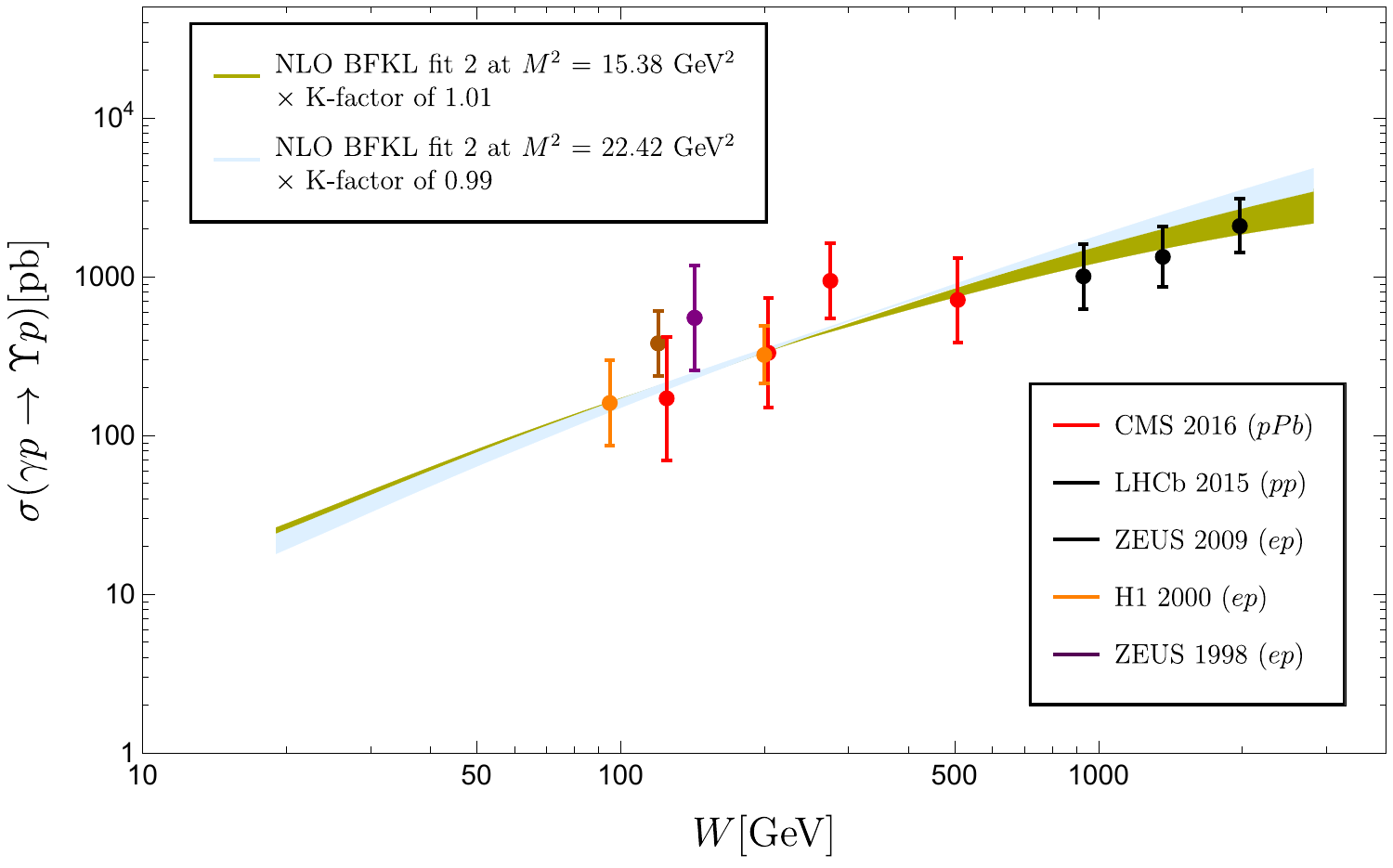}
  \caption{\it Energy dependence of the $\Upsilon$ photo-production
    cross-section as provided by the BFKL fit 1 (up) and 2 (down).
    The uncertainty band reflects a variation of the scale
    $\overline{M}^2 \to \{\overline{M}^2/2, \overline{M}^2 \cdot 2\}$.
    We also show HERA data measured by H1 \cite{Adloff:2000vm} and
    ZEUS \cite{Breitweg:1998ki,Chekanov:2009zz} and LHC data by LHCb
    \cite{Aaij:2015kea} and CMS \cite{CMS:2016nct}.}
  \label{fig:resultsUpsilon}
\end{figure}

Our results for the $W$-dependence of the total $\gamma p \to V p$
cross-section are depicted in Fig.~\ref{fig:resultsJPsi} ($J/\Psi$)
and Fig.~\ref{fig:resultsUpsilon} ($\Upsilon$)and compared to data
from HERA and LHC experiments. Both fits of free parameters of the
proton impact factor, summarized in Tab.~\ref{tab:fit}, are shown in
the plots. We further show results for two different choices of the
`hard' scale $M^2$ of the unintegrated gluon density, {\it i.e.}  the
photoproduction scale $M^2 = M_V^2/4$ and the scale
$M^2 = 8/\mathcal{R}^2$, see Tab.~\ref{vm_fit} for numerical values.
The choice $M^2 = 8/\mathcal{R}^2$ is motivated by the structure of
the impact factor Eq.~\eqref{eq:24}, where it cancels the
($z$-independent part) of the factor $(..)^\gamma$ and therefore
removes the scale dependence\footnote{The impact factor is of course
  still dependent on ratios of other scales. This is natural since,
  unlike the inclusive DIS impact factor, it is not characterized by a
  single hard scale.}; we further find that the this choice seems to
minimize the size of the term $d \ln \Phi_{V,T}/d \gamma$ in
Eq.~\eqref{eq:diffXsec_final}.  We find that our result is only mildly
dependent on this choice. Since the effective Pomeron intercept
increases with increasing hard scale, see also
\cite{Hentschinski:2012kr,Hentschinski:2013id}, the observed rise is
always slightly stronger for the larger of the two scales.  We further
identify $\overline{M} = M$, while we vary $\overline{M}^2$ in the
interval $[M^2/2, 2 M^2]$ to assess the uncertainty associated with
this choice. We find that the result is
rather stable under this variation.\\

Comparing our results with data we find that the overall normalization
obtained from the combination of BFKL gluon density and $\gamma \to V$
impact factor does -- for the majority of cases -- not coincide with
measured data. This is in particular true for the BFKL fit 1, where
typical values of necessary K-factors lie in the range $3 - 3.5$.  The
BFKL fit 2 requires on the other hand only a small ($0.80 - 0.81$ for
$J/\Psi$) or no correction ($0.99 - 1.01$ for $\Upsilon$).  In the
current analysis we fix this normalization by the central values of
some arbitrarily picked low energy data points, {\it i.e.} low energy
ALICE ($J/\Psi$) and ZEUS ($\Upsilon$) data. While it is possible to
improve further the description through fitting the normalization to
the entire data set, we believe that the current treatment is best
suited to study the description of the $W$-dependence, on which we
focus here.\\

Turning to the $W$-dependence we find that both fits and both scale
choices allow for an excellent description of data in the case of
$\Upsilon$-production, see Fig.~\ref{fig:resultsUpsilon}, where fit 2
essentially requires no K-factor. For the $J/\Psi$ data set we find
that fit 2 provides a very good description of the data (with a
$K$-factor $\sim 0.8$ of order one), revealing a slight preference for
the photoproduction scale $M^2 = M_{J/\Psi}^2/4$. The BFKL description
based on the fit 1 also allows for a very good description of the
$W$-dependence up to the last 2 LHCb data points, for which the
predicted growth with $W$ is too slow. Despite of this slight mismatch
of fit 1 in the case of $J/\Psi$ production, we find that the observed
agreement with data is remarkable. This is in particular true for data
points with $W > 500$ GeV which require $x$-evolution beyond the
region constrained by the fit to HERA data and for which the obtained
description directly tests the validity of the present implementation
of NLO BFKL evolution.\\

While the observed mismatch in the overall normalization is not
completely satisfactory, it is somehow expected and -- at least for
the BFKL fit 2 where the correction is small -- easily explained by
the limitations of the current framework.  In the case of fit 1 a
first improvement is obtained if corrections due to $x \neq x'$ (as
available for the collinear gluon distribution function as discussed
in Sec.~\ref{sec:scatt-ampl-high}) are included. Nevertheless also
these corrections are not capable to account for the complete
$K$-factor.  For fit 2 one has to take into account that this fit is
based on a leading order virtual photon impact factor with kinematic
improvements \cite{Kwiecinski:1997ee}, while the currently used impact
factor for the transition $\gamma \to V$ does not contain such
kinematic improvements; in the case of $\gamma \to V$ such corrections
would also include corrections due to $x \neq x'$.  While the
kinematic improvements reduce in the case of DIS studies the magnitude
of the impact factors, corrections due to $x \neq x'$ are for the case
of vector mesons known to enhance the impact factor, at least in the
collinear limit. In the case of fit 2 we therefore expect to a large
extend a cancellation of both effects.  A second point which applies
both to fit 1 and fit 2 is concerned with the treatment of heavy quark
masses: while the impact factor Eq.~\eqref{eq:24} obviously depends on
the heavy quark mass, the original DIS fits are limited to $n_f=4$
mass-less flavors.  Altogether we believe that it is more than
plausible that such effects can account for the observed mismatch in
normalization, in particular in the case of fit 2 where the mismatch
is rather mild.

\section{Outlook and Conclusions}
\label{sec:disc-concl}
In this work we applied the inclusive BFKL fit of
\cite{Hentschinski:2013id} to the description of exclusive vector
meson photo-production at HERA and the LHC. As a new result we
calculated the impact factor for the $\gamma \to V$ transition in the
$\gamma$-Mellin space representation, using earlier result based on
the light-front wave function of vector mesons used in the combination
with color dipole models. Our phenomenological studies show that the
BFKL fits of \cite{Hentschinski:2013id} can provide a very good
description of the center-of-mass energy dependence of the
$\gamma p \to J/\Psi p$ and $\gamma p \to \Upsilon p$ cross-sections.
While the BFKL fit 1 requires a relatively large adjustment in the
overall normalization (of order $3-3.5$), the necessary adjustment is
of order one in the case of BFKL fit 2.  We stress that the current
analysis uses only a fit of the transverse momentum distribution in
the proton, while the $W$-dependence directly results from NLO BFKL
resummation, together with a resummation of collinearly enhanced terms
within the NLO kernel and a optimal renormalization scale setting for
the scale invariant terms of the NLO BFKL kernel. The study provides
therefore direct evidence for the validity of BFKL
evolution at the LHC. \\

Despite of the success of the current description, there are a number
of directions in which our analysis could and should be re-fined. This
implies at first the determination of kinematic corrections to the
impact factor for the transition $\gamma \to V$, which might provide
an opportunity to improve on the observed mismatch in the overall
normalization.  To improve the description further, it will be
necessary to provide a re-fit of HERA data which takes into account
heavy quark masses and possibly now available next-to-leading order
corrections to the virtual photon impact factor with massless quarks.
On the level of the $\gamma p \to V p$ cross-section this would then
further require the determination of corresponding NLO corrections for
the $\gamma \to V$ impact factor, {\it e.g.} using the calculational
techniques developed and used in NLO calculations within high energy
factorization \cite{Hentschinski:2011tz, Hentschinski:2014lma,
  Ayala:2016lhd}.

\subsubsection*{Acknowledgments}
The authors acknowledge support by CONACyT-Mexico grant number
CB-2014-241408. We further would like to thank Laurent Favart for
pointing out an erroneous H1 data point in an earlier version of this paper.

\appendix

\section{Integrals used in the calculation of the impact factor}
\label{sec:integr-used-calc}

To determine
$N({\bm r}, x)$ from the BFKL gluon density, it is necessary to
calculate
\begin{align}
  \label{eq:fromkTtor}
  \int \frac{d^2 {\bm k}}{{\bm k}^2} \left( 1 - e^{i {\bm k} \cdot
  {\bm r}}\right) \frac{1}{({\bm k}^2)^{1-\gamma}} \, .
\end{align}
With $\Re\text{e} \gamma = 1/2$ the individual integrals are not convergent. It is therefore necessary to introduce a a regulator ${\bm k}^2 > \Lambda$ which will set to zero at the end of the calculation (after cancellation of the divergence in Eq.~\eqref{eq:fromkTtor}). We obtain
\begin{align}
  \label{eq:11}
 \lim_{\Lambda \to 0}   \int \frac{d^2 {\bm k}}{\pi} \frac{e^{i {\bm k}\cdot {\bm r}}}{({
\bm k}^2)^{2 - \gamma }} \Theta({\bm k}^2 - \Lambda^2)
 & = \frac{\Gamma(\gamma -1)}{\Gamma(2 - \gamma)} \left(\frac{{\bm r}^2}{4} \right)^{1-\gamma} + \frac{\Lambda^{\gamma-1}}{1-\gamma}\, ,
\end{align}
while
\begin{align}
  \label{eq:15}
   \int \frac{d^2 {\bm k}}{\pi} \frac{ \Theta({\bm k}^2 - \Lambda)}{({
\bm k}^2)^{2 - \gamma}} & = \frac{\Lambda^{\gamma -1}}{1-\gamma}.
\end{align}
and therefore
\begin{align}
  \label{eq:fromkTtorFERTIG}
  \int \frac{d^2 {\bm k}}{{\bm k}^2} \left( 1 - e^{i {\bm k} \cdot
  {\bm r}}\right) \frac{1}{({\bm k}^2)^{1-\gamma}}  &=
-\pi   \frac{\Gamma(\gamma -1)}{\Gamma(2 - \gamma)} \left(\frac{{\bm r}^2}{4} \right)^{1-\gamma}  = \frac{  \pi\Gamma(\gamma) 4^{\gamma-1}}{(1-\gamma) \Gamma(2-\gamma) ({\bm r}^2)^{\gamma-1} } \, .
\end{align}
In a second step we need to integrate over the dipole size $r$. With
\begin{align}
  \label{eq:22}
  K_0(r \epsilon) & =
\frac{1}{2} \int_0^\infty \frac{d \lambda}{\lambda} e^{-\lambda \epsilon^2 - \frac{{\bm r}^2}{4 \lambda}},
&
   \frac{r}{\epsilon} K_1(r \epsilon)  & =
\frac{1}{2} \int_0^\infty {d \lambda} e^{-\lambda \epsilon^2 - \frac{{\bm r}^2}{4 \lambda}},
\end{align}
this can be done using the following integral:
\begin{align}
  \label{eq:12}
  f(a,b, Q_0^2) & =  \int_0^\infty {d \lambda} \lambda^{a-1}  \int \frac{d^2 {\bm r}}{\pi}    e^{-\lambda \epsilon^2 - \frac{{\bm r}^2}{4 \lambda}} e^{- {\bm r}^2 Q_0^2} \frac{1}{({\bm r}^2)^b} \notag \\
&=
\Gamma(1-b) \Gamma(1+a-b) 4^{-a} \left( Q_0^2 \right)^{b-a-1} U\left(1+a-b, 1+a, \frac{\epsilon^2}{4 Q_0^2} \right)
\end{align}
where $U$ is a Hypergeometric function of the second kind or Kummer's
function.

\end{document}